\documentclass[11pt]{article}
\usepackage[]{psfig}
\newcommand{\be}{\begin{equation}}
\newcommand{\ee}{\end{equation}}
\newcommand{\ba}{\begin{eqnarray}}
\newcommand{\ea}{\end{eqnarray}}

\begin{document}
\title{\bf A monopole solution from noncommutative multi-instantons}
\author{
D.H.~Correa$^a$\thanks{CONICET} \, , P.~Forg\'acs$^{b}$ \,, E.F.~Moreno$^{a,c\, *}$ \\
F.A.~Schaposnik$^a$\thanks{Associated with CICPBA}
and G.A.~Silva$^{a *}$  \\
{\normalsize\it  $a$ Departamento de F\'\i sica, Universidad Nacional
de La Plata}\\
{\normalsize\it C.C. 67, 1900 La Plata, Argentina}
\\
{\normalsize\it $^b$Laboratoire de Math\'ematiques et Physique Th\'eorique}\\
{\normalsize\it CNRS/UMR 6083, Universit\'e de Tours}\\
{\normalsize\it Parc de Grandmont, 37200 Tours, France}
\\
{\normalsize\it $^c$Department of Physics,West Virginia University}\\
{\normalsize\it Morgantown, West Virginia 26506-6315, U.S.A.}
}

\date{\today}
\maketitle

%===================================================================
%===================================================================
\begin{abstract}
We extend the relation between instanton and monopole solutions of
the selfduality equations in $SU(2)$ gauge theory to
noncommutative space-times. Using this approach and starting from
a noncommutative multi-instanton solution we construct a $U(2)$
monopole configuration which lives in 3 dimensional ordinary
space. This configuration resembles the Wu-Yang monopole and
satisfies the selfduality (Bogomol'nyi) equations for a $U(2)$
Yang-Mills-Higgs system.
\end{abstract}

%\date{\today}

\section{Introduction}
Soliton and instanton solutions in noncommutative field theories
have been the object of many investigations in recent years (see
\cite{Douglas}-\cite{Szabo} for a complete list of references).
Whenever the space-time dimension is even, a connection between
the noncommutative algebra and that of creation and annihilation
operators in Fock space can be exploited in order to find explicit
exact solutions which are the natural extensions of those already
constructed in ordinary space.  In particular, after the
pioneering work of Nekrasov and Schwarz \cite{NS} on
noncommutative instantons, different approaches have been followed
to construct and analyze explicit selfdual solutions
\cite{ins1}-\cite{insN}.

Concerning solitons, and apart from vortices (see \cite{fs} for a
complete list of references on this issue), noncommutative monopole
configurations have been extensively discussed
\cite{mon1}-\cite{monN}. In particular, BPS  monopole solutions
have been constructed in \cite{Gross3} for  noncommutative $U(1)$
and $U(2)$ gauge theories by solving the noncommutative extension
of the so-called Nahm equations. An interesting correspondence
between the noncommutative monopole solution and a D1 string
stretched between D3 branes was revealed by this work.

As it is well known, conventional instanton and monopole solutions
can be related. Geometrically, the idea is that if one looks for
solutions of the selfduality equations with a $U(1)$ isometry
$k^\mu$, then a monopole configuration of a Yang-Mills-Higgs
system  can be obtained with $\Phi = k^\mu A_\mu$ playing the role
of the Higgs scalar in the adjoint. When the isometry is chosen to
be along the Euclidean time ($\Phi =A_0$) the selfduality
equations become the Bogomol'nyi equations for a Yang-Mills-Higgs
system in the Prasad-Sommerfield limit. This procedure, originally
developed in Refs. \cite{manton}, starting from an axially
symmetric multi-instanton solution with charge $q$ \cite{witten},
was afterwards extended by Nahm \cite{nahm}  to the ADHM
multi-instanton solution. A different choice for $k^\mu$ leading
to hyperbolic monopoles was originally proposed by Atiyah
\cite{atiyah}.

The extension of Nahm's construction to the noncommutative case
has been developed in \cite{Gross1}-\cite{Gross3}, based on the
noncommutative version of the ADHM construction developed in
\cite{NS}. In the $U(1)$ case, which was studied in detail in
\cite{Gross1}, a soliton solution  having  zero magnetic charge
was constructed. It can be interpreted as consisting of a monopole
attached to a string that runs off to infinity. In order to see
whether truly magnetically charged isolated configurations in
3-dimensional noncommutative space can be obtained from
4-dimensional noncommutative instantons we shall extend in this
work Manton's proposal of considering the infinite charge limit
($q \to \infty$) of Witten multi-instanton  solution. We will use
the noncommutative version of Witten's solution constructed in
\cite{Correa2}, which we review  in section 2. Then, in section 3
we discuss the choice of the appropriate  gauge condition and
discover, as a byproduct, a very peculiar situation that can arise
for constant field strengths in noncommutative gauge theories.
Indeed, we show that under certain conditions, there exist gauge
orbits consisting of just one point. The $q \to \infty$ limit
leading to a monopole configuration is considered in section 4
where we write the BPS equations obeyed by the soliton solution.
We discuss the properties of the solution, relating it with that
of a Dirac monopole. Finally, in section 5 we summarize and
discuss our results.

\section{The instanton solution}
We here briefly review the extension of Witten's multi-instanton
solution to noncommutative space, as presented in
ref.\cite{Correa2}.

The clue in Witten's ansatz  \cite{witten} is to reduce the four
dimensional problem to a two dimensional one through an axially
symmetric multi-instanton ansatz. That is, one passes from 4
dimensional Euclidean space-time with coordinates
($r,\vartheta,\varphi,t $) to 2 dimensional curved space-time with
coordinates $(r,t)$.

The noncommutative solution in \cite{Correa2} corresponds to a
space-time with commutation relations given by
\begin{eqnarray}
&& [r,t] = i\theta(r,t)\nonumber\\
&& [r,\vartheta] = [r,\varphi]=[t,\vartheta] = [t,\varphi] =
[\vartheta,\varphi] = 0
\label{tehtas}
\end{eqnarray}

Eq.(\ref{tehtas}) corresponds to the most natural commutation
relations to impose when a problem with cylindrical symmetry is to
be studied. In principle, $\theta(r,t)$ in (\ref{tehtas}) is  an
arbitrary function. However, noncommutativity in curved space-time
imposes severe restrictions on the function $\theta(r,t)$. In
general, given a two-dimensional space-time with coordinates
$x^i$, $i=1,2$ and commutation relations of the general form
\be
[x^i,x^j] = i \theta^{ij}(x)\;\; ,
\label{sss}
\ee
the associativity of the product is not guaranteed for an
arbitrary function $\theta^{ij}(x)$. One can see however that
associativity can be achieved whenever
\be
\nabla_k \theta^{ij} = 0 \; .
\ee
The unique solution of these equations for $d=2$ is  given by
\be
\theta^{ij} = \theta_0 \frac{\varepsilon^{ij}}{\sqrt g}
\label{solucioni}
\ee
with $\theta_0$ being a constant.

The two-dimensional curved space-time metric in which the original
4-dimensional Yang-Mills action reduces to an Abelian Higgs action
turns out to be
\be g^{ij} = r^2 \delta^{ij} \; ,
\label{metrica} \ee
Of course to exploit this connection one necessarily has to
interpret $r$ as a dimensionless variable. This can be achieved by
starting from  dimensionless variables in Euclidean four
dimensional  space (through the introduction of a length scale $R$
which can be related with the instanton size). Alternatively one
can introduce a dimensionful noncommutative parameter $\theta
={R}^2\theta_0$ .

Then, using solution (\ref{solucioni}), we see that the
commutation relations (\ref{tehtas}) to impose should take the
form
\be
[r,t] = i r^2 \theta_0 \; ; \;\;\; {\rm all~other} ~ [.,.]=0
\label{rara}
\ee
with $r$ and $t$  dimensionless variables in the two-dimensional
curved space. The connection with dimensionful variables $r'$ and
$t'$ goes as follows. The two-dimensional curved metric
(\ref{metrica}) should be written in the form
\be
g^{ij} = \frac{{r'}^2}{R^2} \delta^{ij}\label{metrica'}
\ee
so that the commutation rule (\ref{rara}) becomes
\be
[r',t'] = i  {r'}^2 \theta_0
%= i  {{r'}^2} \frac{\theta}{R^2}
\label{rara'}
\ee
 One can easily show that this commutation rule coincides with
that studied in \cite{mp}. From here on we shall work with
dimensionless variables and recover the scale at the end of the
calculations.

A  simplification occurs after the observation that
\be
 r*t - t* r = ir^2 \theta_0  ~ ~ ~
\Rightarrow ~ ~ ~ t   * \frac{1}{r} - \frac{1}{r} * t = i\theta_0
\ee
Then, introducing $y^1 = -1/r$ and $y^2 = t  $ Eq.\ (\ref{rara})
becomes a usual two-dimensional Moyal product,
\be
[y^1,y^2] = i\theta_0
\label{moyali}
\ee
Axially symmetric multi-instanton solutions to the selfduality equations
\be
F_{\mu\nu} = \pm {\tilde{F}}_{\mu \nu}
\label{sd}
\ee
were found in \cite{Correa2} by making a noncommutative extension
(with $U(2)$ gauge group) of the cylindrically symmetric  ansatz
considered by Witten \cite{witten}. For the $SU(2)$ sector one
just proposes the same ansatz as in ordinary space,
\begin{eqnarray}
\vec{A}_{1}  & = &  A_{1}(y^1,y^2) \vec \Omega(\vartheta,\varphi)  \nonumber\\
\vec{A}_2  & = &  A_2(y^1,y^2) \vec \Omega(\vartheta,\varphi) \label{dividido}\\
\vec A_\vartheta &=& \phi^1(y^1,y^2) \partial_\vartheta \vec
\Omega(\vartheta,\varphi) +
\left(1 + \phi^2(y^1,y^2)\right) \vec \Omega (\vartheta,\varphi)
 \wedge \partial_\vartheta \vec \Omega (\vartheta,\varphi)
\nonumber\\
\vec A_\varphi &=& \phi^1(y^1,y^2)
 \partial_\varphi \vec \Omega(\vartheta,\varphi)
 +
\left(1 + \phi^2(y^1,y^2)\right) \vec
\Omega (\vartheta,\varphi) \wedge \partial_\varphi \vec \Omega
(\vartheta,\varphi)\nonumber\\
\label{ansatz}
\end{eqnarray}
with
\begin{equation}
\vec \Omega(\vartheta, \varphi)  = \left(
\begin{array}{c}
\sin \vartheta \cos \varphi \\
\sin \vartheta \sin \varphi \\ \cos \vartheta
\end{array}
\right)
\ee
Concerning the  remaining $U(1)$ components, it is natural to
propose the ansatz
\begin{eqnarray}
A_1^4 &=& A_1^4(y^1,y^2)  \nonumber\\
A_2^4 &=&  A_2^4(y^1,y^2) \nonumber\\
A_\vartheta^4 \!\! &=&\!\! A_\varphi^4 = 0
\label{cuatro}
\end{eqnarray}
With this ansatz, the selfduality equations (\ref{sd}) become
\begin{eqnarray}
&&\!\!\!\!\!\!\!\!\partial_2 A_1 - \partial_1 A_2 + \frac{i}{2} [A_2,A_1^4] + \frac{i}{2}
[A^4_2,A_1] = 1 - \left(\phi^1\right)^2 - \left(\phi^2\right)^2
\nonumber\\
&&\!\!\!\!\!\!\!\!\partial_2 A_1^4 - \partial_1 A_2^4 + \frac{i}{2} [A_2^4,A_1^4]
+ \frac{i}{2} [A_2,A_1] =  -i[\phi^1,\phi^2]
\nonumber\\
&&\!\!\!\!\!\!\!\!\partial_2 \phi^1 + \frac{1}{2} [A_2,\phi^2]_+ + \frac{i}{2} [A_2^4,\phi^1]
= \left(y^1\right)^2\left( \partial_1\phi^2 - \frac{1}{2} [A_1,\phi^1]_+ +
 \frac{i}{2}
[A_1^4,\phi^2]
\right)
\nonumber\\
&&\!\!\!\!\!\!\!\!\partial_2 \phi^2 - \frac{1}{2} [A_2,\phi^1]_+ + \frac{i}{2} [A_2^4,\phi^2]
= -\left(y^1\right)^2\left( \partial_1\phi^1 + \frac{1}{2} [A_1,\phi^2]_+ +
\frac{i}{2}
[A_1^4,\phi^1]
\right) \nonumber\\
\label{largas}
\end{eqnarray}
Imposing the further restriction in the $U(1)$ sector,
\begin{eqnarray}
A_t^4(u,t) = A_t(u,t) \nonumber
\\
A_u^4(u,t) = A_u(u,t) \label{simpl}
\end{eqnarray}
and introducing the notation
\begin{eqnarray}
\phi &=& \phi^1 - i \phi^2 \nonumber
\\
D\phi &=& \partial \phi +i A\phi \nonumber\\
F_{12} &=& \partial_1 A_2 - \partial_2 A_1 + i[A_1,A_2]
\label{tres}
\end{eqnarray}
the system (\ref{largas}) reduces to
\begin{eqnarray}
F_{12} &=& \frac{1}{2}[\phi,\bar \phi] \label{suno}\\
F_{12} &=& \frac{1}{2}[\phi,\bar \phi]_+ -1 \label{sdos}\\
D_2\phi &=& i \left(y^1\right)^2 D_1\phi\
\label{cortas}
\end{eqnarray}
Although this system is overconstrained, nontrivial solutions were
obtained in \cite{Correa2} within the Fock space framework. In this
approach, the noncommutative coordinates algebra defined by
(\ref{moyali}) is viewed as an algebra of annihilation and
creation operators,
\ba a =
\frac{1}{\sqrt{2\theta_0}}\left(y^1 + i y^2\right) \; , && a^\dagger =
\frac{1}{\sqrt{2\theta_0}}\left(y^1 - i y^2\right)
\nonumber\\
~[\, a , a^\dagger ] &=& 1
\label{dager}
\ea
Given a  field $\chi$, one associates an operator $O_\chi$ acting
on Fock space as
\be
O_\chi(a,a^\dagger)  =
\frac{1}{4\pi^2 \theta_0}
\int d^2k
\tilde\chi(k,\bar k)
\exp\left(-i\left(\bar k a + k a^\dagger\right)\right)
\label{adaga}
\ee
The star product of fields in configuration space becomes just the
operator product in Fock space.
\be
O_\eta O_\chi  = O_ {\eta*\chi}
\ee
Here the Moyal $*$-product of two functions $\eta$ and $\chi$ is
defined as
\be
\eta(x)*\chi(x)= \left. \exp\left (\frac{i}{2}
\theta^{ij}\partial_i^x\partial_j^y\right ) \eta(x)\chi(y)
\right\vert_{y=x}
\ee
Derivatives in configuration space should be replaced by
commutators in Fock space,
\be
\partial_z \to  -\frac{1}{\sqrt{\theta_0}} [ a^\dagger,~]
\, , \;\;\;\; \partial_{\bar z} \to   \frac{1}{\sqrt{\theta_0}} [a,~]
\label{conmutads}
\ee
where we have written
\be
z = \frac{1}{\sqrt 2} \left(y^1 + i y^2\right)
\label{zeta}
\ee

Now, compatibility of equations (\ref{suno}) and (\ref{sdos}) implies
\be
\bar \phi \phi = 1 \; , \;\;\; \phi \bar \phi = 1 + 2 F_{12}
\ee
and hence a nontrivial solution exists in the form of a shift
operator,
\be
\phi = \sum_{n=0} |n + q\rangle \langle   n|
\label{silva}
\ee
Here $\{|n\rangle\}$ is the Fock space basis of eigenfunctions of
the number operator $ N = a^\dagger a$ and the integer $q \geq 0$
is related to the topological charge. Now, consistency of this
last equation with eq.(\ref{cortas}) completely fixes $A_z$,
\begin{eqnarray}
A_z &=& -\frac{i}{\sqrt{\theta_0}} \sum_{n=0}^{q-1} \left( \sqrt{n+1}
\right) |n+1\rangle\langle n|\label{solucion2} +\nonumber\\
&& ~ ~ ~ +\frac{i}{\sqrt{\theta_0}} \sum_{n=q} \left( \sqrt{n+1-q} - \sqrt{n+1}
\right) |n+1\rangle\langle n|
\label{solucion}
\end{eqnarray}
provided that
\be \theta_0 = 2 \label{dosss}\ee
In particular, both the l.h.s. and r.h.s of eq.(\ref{cortas})
vanish separately. Regarding the particular value of $\theta_0$
for which  the solution was found, let us recall that also for
vortices in  flat space it was necessary to fix $\theta_0$ (but in
that case to the value $\theta_0 = 1$),  in order to satisfy the
corresponding Bogomol'nyi equations.

The magnetic field $B = i F_{z\bar z}$ associated with solution
(\ref{solucion2}) takes the form,
\be
B = -\frac{1}{2} \left(|0\rangle\langle 0|+ ...
+ |q-1\rangle\langle q-1|\right)
\label{bb}
\ee
with associated magnetic flux
\be
\Phi = 2\pi   Tr B  = -{\pi} q
\ee
A factor $\pi \theta_0$ was included in the definition of the
magnetic flux, one half of the usual factor  since one is working
in the half plane.

Each projector $|n\rangle \langle n|$ in Fock space can be related
to a Laguerre polynomial in configuration space through the
connection
\be
|n\rangle \langle n|  \to 2 (-1)^n  \exp\left(-\frac{(y^1)^2
+ (y^2)^2}
{2}\right)L_n\left((y^1)^2
+ (y^2)^2\right)
\label{lag}
\ee
Then, since the Laguerre polynomial $L_n$ is concentrated in an
annulus of radius $R_n$, growing with $n$ according to  $R_n\sim
\sqrt n$, one can view the magnetic flux (\ref{bb}) as that of a
superposition of $q$ annular vortices of unit flux. This should be
compared with the multi-instanton solution in ordinary space, for
which the corresponding $q$-vortex is a superposition of $q$
1-vortices centered at arbitrary points along the time axis.

We can now easily write  the selfdual multi-instanton solution in
$4$-dimen\-sio\-nal space by inserting the solution
(\ref{silva})-(\ref{solucion}) into the ansatz (\ref{ansatz}). The
resulting selfdual field strength reads
\begin{eqnarray}
\vec F_{21} &=&   B \vec \Omega \label{instanI}\\
\vec F_{\vartheta\varphi}
&=& B   \sin \vartheta \, \vec \Omega \label{instanII}\\
F_{21}^4 &=&    B \label{instanIIII}\\
 F_{\vartheta\varphi}^4 &=& B \sin \vartheta
\label{instanIV}
\end{eqnarray}
with the other field-strength components vanishing.
The instanton number is given by
\begin{equation}
Q = \frac{1}{32 \pi^2} {\rm tr} \int d^4 x {\varepsilon}^{\mu \nu \alpha \beta}
 F_{\mu\nu} F_{\alpha \beta}
= \frac{1}{\pi}
\int_{-\infty}^{0}dy^1 \int_{-\infty}^{\infty} dy^2
 B^2 = 2\, {\rm Tr} B^2
= \frac{q}{2}
\end{equation}

\section{Gauge choices}
As stated in the introduction,  Manton \cite{manton} developed a
procedure (that implies taking the limit of infinite topological
charge) that effectively reduces  the 4 dimensional cylindrically
symmetric multi-instanton configuration in ordinary space  to a
static monopole solution of the Bogomol'nyi-Prasad-Sommerfield
equations. In order to extend this procedure to the noncommutative
case, we shall need to consider the instanton configuration
described in the precedent section in an appropriate gauge
ensuring that, after taking the $q \to \infty$ limit, one ends,
after an appropriate time-dependent gauge transformation,  with a
static configuration so that the remaining spatial dependence will
be consistent with static BPS equations of a Yang-Mills-Higgs system.

Now, as we shall see, after taking the $q \to \infty$ limit of the
noncommutative instanton described above, the gauge field
configuration, as it happens in the commutative case, remains time
dependent. This is due to the fact that the 2-dimensional vortex
solution from which it was constructed, originally in the Lorentz
gauge, becomes, in the infinite charge limit, a linear function
with one  of its components  depending on $t$. In ordinary space,
such a linear  dependence  on time can be easily eliminated by an
appropriate  gauge transformation but the procedure becomes
delicate in the noncommutative case. We shall then discuss this
point (at the level of the vortex solution), before proceeding to
the analysis of the resulting BPS equations.

Let us consider a $U_*(1)$ linear gauge potential in $d=2$
dimensions, in the Lorentz gauge,
\be
{\cal A}_i = \frac{\cal B}{2} \varepsilon_{ij} x^j \; , \;\;\; i,j=1,2
\label{exp}
\ee
where the commutation relations for coordinates are
\be
[x^1,x^2] = i \theta_0
\ee
The field strength takes the form
\begin{eqnarray}
F_{12} &=& \partial_1 {\cal A}_2 - \partial_2 {\cal A}_1
+ i \left(
{\cal A}_1 * {\cal A}_2 - {\cal A}_2 * {\cal A}_1\right) \nonumber\\
&=&
-{\cal B} - \frac{{\cal B}^2\theta_0}{4}
\label{F}
\end{eqnarray}
The first term in the second line of (\ref{F}) is just the field
strength that would arise in the commutative case, while the
second is due to the fact we are dealing with the noncommutative
$U(1)$ gauge group, which we denote by $U_*(1)$. With our
conventions, the covariant derivative in the adjoint reads
\be
D_i = \partial_i + i [{\cal A}_i,~]
\label{cov}
\ee
Considering a gauge transformation under which gauge fields
change as
\begin{eqnarray}
{\cal A}'_i &=& g^{-1}*{\cal A}_i*g  - i g^{-1} * \partial_i g  \label{a1}\\
F'_{ij} &=& g^{-1}*F_{ij} *g \label{f1}
\end{eqnarray}
then, eq.(\ref{a1}) can be written in the form
\be
{\cal A}'_i = {\cal A}_i + g^{-1}*[{\cal A}_i,g] - i g^{-1} * \partial_i g
\label{prima}
\ee
Now, in view of the explicit form of the gauge field configuration
(\ref{exp}) one has
\be [{\cal A}_i,g] = -i\frac{{\cal B}\theta_0}{2}\partial_i g \ee
so that, finally, eq.(\ref{prima}) becomes
\be
{\cal A}'_i = {\cal A}_i -i\alpha   g^{-1} \partial_i g \label{aqui}
\ee
where
\be
\alpha  = 1+\frac{{\cal B}\theta_0}{2}
\ee
We then see that, if one only allows for regular gauge
transformations, the gauge orbit to which ${\cal A}_i$ belongs
consists, for ${\cal B}\theta_0 =- 2$, of just one point. But it
is precisely the value to which our multi-vortex solution tends in
the $q\to\infty$ limit. As we shall show by allowing gauge
transformations singular at ${\cal B}\theta_0 = -2$, one is able
to gauge away the ${\cal A}_1$ component of the configuration
(\ref{exp}).

In the  commutative case, one easily finds that the
transformation corresponds to the gauge group element
\be g_{c} = \exp \left( - i \frac{\cal B}{2} x^1x^2\right) \ee
We then propose the following ansatz for the  gauge transformation
in the noncommutative case,
\be
g_{nc} = A \exp \left( -i  \beta x^1x^2\right)
\label{ge}
\ee
where
\be
\beta = \frac{{\cal B}}{1 +\alpha}\,
\ee
and $A$ is an arbitrary parameter to be appropriately adjusted.
Note that the exponential in (\ref{ge}) is defined with the
ordinary product in its series expansion
\begin{eqnarray}
g_{nc} \!\!\!& =&  \!\!\!A
\left(
1 -i\left (\beta x^1x^2\right)-  \frac{1}{2!}
\left (\beta x^1x^2\right) \left (\beta x^1x^2\right)
+ \ldots
\right) \nonumber\\
\label{ge2}
\end{eqnarray}
Because of this fact, it is not a priory guaranteed that $g_{nc}$
is a unitary element of the noncommutative gauge group $U_*(1)$.
We shall see however that one can chose $A$ so that $g_{nc} \in
U_*(1)$. To see this, it will be convenient to use the Weyl-Moyal
connection (\ref{adaga}),
\be
{\hat g_{nc}}({\hat x}_1,{\hat x}_2) = \int\; \frac{d^2p}{(2\pi)^2} \;
{\tilde g_{nc}}(p)
\;e^{i(p_1{\hat x}_1 + p_2 {\hat x}_2)}
\ee
where ${\hat x}_1$ and ${\hat x}_2$ are operators satisfying the
noncommutative algebra,
\be
[{\hat x}_1,{\hat x}_2]
= i
\theta
\ee
In this framework, the product of operators can be written in
Fourier space as
\be
{\hat f}({\hat x}_1,{\hat x}_2) \cdot {\hat h}({\hat x}_1,{\hat x}_2)
\to
\int\; \frac{d^2q}{(2\pi)^2} \; {\tilde f}(p-q)
{\tilde h}(q) \;
\exp(i(p_1 q_2 - p_2 q_1)\theta_0)
\ee
For the ansatz (\ref{ge}) one has
\be
{{\tilde g}_{nc}}(p_1,p_2) = \frac{2\pi A}{\beta}\;
 \exp\left(i \frac{p_1 p_2}{\beta}\right)
\ee
Then, after some straightforward calculation, one finds
\[
{\hat g}_{nc}({\hat x}_1,{\hat x}_2 ) \cdot {\hat g}_{nc}({\hat x}_1,{\hat x}_2 )
^{\dagger} =
\frac{|A|^2}{1-(\theta_0
\beta/2)^2}
\]
Finally, with an appropriate choice for $A$ one can write the
unitary gauge transformation ${g}_{nc} \in U_*(1)$ in the form
\be
{g}_{nc} =
\frac{\sqrt{1 + {\cal B}\theta_0/2}}{1 + B\theta_0/4}
\exp\left(
-i\frac{\cal B}{2(1 + {\cal B}\theta_0/4)} x_1x_2
\right)
\ee
Under this gauge transformation, which as  expected  is singular
at $\theta_0 {\cal B}= -2$, one manages to gauge out the ${\cal
A}_1$ component in (\ref{exp}),
\be {\cal A}'_1 = g^{-1}_{nc}*{\cal A}_1* g_{nc} - i g^{-1}_{nc}*
\partial_i g_{nc} = 0 \label{alfin} \ee

Let us now uplift this transformation to the full gauge group
$U_*(2)$, in order to eliminate an $A_1$ linear component   in the
original 4 dimensional ansatz (\ref{ansatz}) and (\ref{cuatro}).
We propose the following gauge group transformation
\be
g_{U(2)} = \exp_{*}\left(-ic[y^1,y^2]_+\Lambda\right)
\label{propose}
\ee
with $[y^1,y^2]_+$ the Moyal anticommutator  of $y^1$ and $y^2$ and
\be c = \frac{1}{2\theta_0}\log \left(1 + \frac{{\cal
B}\theta_0}{2} \right) \ee
\be
\Lambda = \frac{1}{2} (\Omega^a \sigma^a + I)
\label{Lambda}
\ee
The notation $\exp_{*}$ means that this exponential is defined using the Moyal product
in its series expansion.

One can easily see that
\be
g_{U(2)}^\dagger =  g_{U(2)}^{-1} = 1
+ \Lambda\left(g_{nc}^\dagger -1\right)
\ee
The $U_*(2)$ gauge transformation for the $i=1,2$ components of the
$A_i$ transform according to
\begin{eqnarray}
A'_i &=& g_{U(2)}^{-1} * A_i * g_{U(2)}  + i g_{U(2)}^{-1} *
\partial_i g_{U(2)}
\nonumber\\
&=& \Lambda \left(g_{nc}^{-1} * A_i * g_{nc} + i g_{nc}^{-1}
*\partial_i g_{nc} \right)
 \; , \;\;\; i=1,2
\end{eqnarray}
so, in view of (\ref{alfin}), one can gauge out the linear time
dependent component $A_1$ of the gauge field configuration leading
to the field strength (\ref{instanIV}). ~

\section{Monopoles from instantons}

Let us now consider the limit of infinite topological charge in
order to construct static, spherically symmetric BPS solutions
from axially symmetric ones. First, taking the $q \to \infty$
limit in Eq.\ (\ref{bb}) one gets a constant  magnetic field,
\be \lim_{q \to \infty} B= -\frac{1}{2} \sum_{n=0}^{\infty}
|n\rangle \langle n| = -\frac{1}{2} \label{unos} \ee
Such a magnetic field follows from the gauge field configuration
(see eq.(\ref{solucion}))
\begin{equation}
\lim_{q \to \infty} A_z =
-\frac{i}{\sqrt{2}} \sum_{n=0}^{\infty} \left( \sqrt{n+1}
\right) |n+1\rangle\langle n| =
-\frac{i}{\sqrt{2}} \sum_{n=0}^{\infty}  a^\dagger |n\rangle\langle n|
= -\frac{i}{2} \bar z
\label{solucion3}
\end{equation}
Recalling that $A_z = (1/\sqrt 2) (A_1 - i A_2)$ we have
\be
\lim_{q \to \infty} A_1 = -\frac{y^2}{2} \; , \;\;\; \;\;\; \;\;\;
\lim_{q \to \infty} A_2 = \frac{y^1}{2}
\label{dependencia}
\ee

In order to convert the instanton selfduality equations (\ref{sd})
into static BPS equations for a Yang-Mills-Higgs system, one first
needs to identify the time component $A_2$ of the gauge field with
a Higgs scalar $\Phi$ taking values in the Lie algebra of
$U_*(2)$. The spatial components $(A_1,A_\vartheta,A_\varphi) $
will be identified with the spatial components  of a Yang-Mills
field that we shall denote $B_i$. That is, taking $B_0=0$ one
establishes the following connection
\begin{eqnarray}
A_2 &\to& \Phi \nonumber\\
(A_r,A_\vartheta,A_\varphi) &\to& (B_r,B_\vartheta,B_\varphi)
\nonumber\\
F_{ij} &\to & G_{ij} = \partial_i B_j - \partial_j B_i +
i[B_i,B_j] \label{Higgs}
\end{eqnarray}
Now, in order to obtain a noncommutative  $U_*(2)$ monopole like
static solution $(B_i,\Phi)$ from the instanton solution $A_\mu$
as defined in (\ref{dividido})-(\ref{ansatz}) one needs a
time-independent field configuration. While  the $q \to \infty$
limit does lead to a static configuration for the Higgs field
$\Phi$, this is not the case for the gauge field components. The
$A_1$ component exhibits a linear dependence on $y^2 = t$, as
given by eq.(\ref{dependencia}), which could be gauged away, but
subject to a proviso related  to the discussion   in section 3.
Indeed, we have seen that a two-dimensional configuration of  the
type (\ref{exp}), with ${\cal B}=-1$ (or ${B}\theta_0 = -2$)
exhibits a gauge orbit consisting of just one point and the same
happens for our 4-dimensional $U_*(2)$ configuration.  Then, to
gauge away the $y^2$ (time) dependence of  $A_1$ we are forced to
consider singular gauge transformations of the kind discussed in
section 3. Indeed, under a gauge transformation of the form
(\ref{propose})
\be
g_{U(2)} = \exp_{*}\left(-ic[y^1,y^2]_+\Lambda\right)
\label{proposea}
\ee
$A_1$ vanishes while $A_2$ becomes
\be
A_2 = -{\cal B}\left(1 +  \frac{{\cal B}\theta_0}{4} \right)  x^1\Lambda =
\frac{1}{2}
x^1 \Lambda
\label{sushi}
\ee
Then,  the $U_*(2)$  Higgs scalar $ \Phi = A_2 $ is just
\be
\Phi = \frac{1}{2} x^1 \Lambda  = -\frac{1}{2r}  \Lambda
\label{esca}
\ee

Finding the actual gauge transformation that eliminates the time
dependence from the angular components is far more complicated.
However,  we know that the in the $q \to \infty$ limit the only
non-trivial strength components of the gauge field, as given by
eqs. (\ref{instanI})-(\ref{instanIV}) take the very simple form
\begin{eqnarray}
&&\vec F_{0r} = \frac{B}{r^2} \, \vec \Omega \; ,\;\;\;\; \vec F_{\vartheta\varphi}
= B   \sin \vartheta \, \vec \Omega\\
&&F_{0r}^4 = \frac{B}{r^2} \; , \;\;\;\;\;\; F_{\vartheta\varphi}^4 =
 B \sin \vartheta
 \label{instanIVq}
\end{eqnarray}
%
%\begin{eqnarray}
%\vec G_{\vartheta\varphi}
%&=& B  \sin \vartheta \, \vec \Omega \nonumber\\
% G_{\vartheta\varphi}^4 &=& B \sin \vartheta
%
%\end{eqnarray}
%%
with $B = -1/2$. One can then easily find a time-independent
instanton configuration leading to such a field strength. It is
simply given by
\begin{eqnarray}
&&\vec A'_0 = \frac{B}{r}\, \vec \Omega \; , \;\;\;
\vec A'_r   = 0\; , \;\; \vec A'_\vartheta = -  \vec \Omega
\wedge \partial_\vartheta \vec \Omega \; , \nonumber \\
&&\vec A'_\varphi = - \vec \Omega \wedge \partial_\varphi\vec \Omega
- (B + 1) (1 + \cos \vartheta) \vec \Omega \cr
&&{A'}_0^4 = \frac{B}{r} \; , \;\; {A'}_r^4  = 0  \; , \;\;
{A'}_\vartheta^4 = 0  \; , \;\; {A'}_\varphi^4 = - B(1 + \cos
\vartheta) \label{dirit}
\end{eqnarray}
Since for nonabelian gauge theories the field strength does not
determine the gauge potential up to gauge transformations, as was
shown by Wu and Yang in his classic article \cite{wu-yang}, is not
obvious that the fields $A'_{\mu}$ in (\ref{dirit}) are gauge
equivalent to the original instanton configuration $A_{\mu}$.
However we will show that this is in fact the case, the gauge
configurations $A'_{\mu}$ and $A_{\mu}$ are related by a gauge
transformation.

To see this we notice that both gauge configurations generate the
same field strength and satisfy the same equations of motion.
Concerning the Bianchi identities, they are both satisfied
everywhere except at the origin where they both have the same
delta function singularity (see the discussion below). Most of the
components of $F_{\mu \nu}$ vanishes, so that from the equation of
motion we deduce the following identities
\begin{eqnarray}
D_0 F_{0r}=0 \; , \;\;\; &&D_r F_{0 r}=-\frac{2}{r}\, F_{0r} \cr
D_{\vartheta} F_{\vartheta \varphi}=0 \; , \;\;\; && D_{\varphi} F_{\vartheta
\varphi}=0
\end{eqnarray}
and from the Bianchi identities
\begin{eqnarray}
D_{\vartheta} F_{0r}=0 \; , \;\;\; &&D_{\varphi} F_{0 r}= 0 \cr
D_{0} F_{\vartheta \varphi}=0 \; , \;\;\; && D_{r} F_{\vartheta \varphi}=
2\pi\delta^{(3)}  \Lambda
\label{delta}
\end{eqnarray}
Then we see that all the covariant derivatives of $F_{\mu \nu}$
vanishes, except for $D_r F_{0 r}=-\frac{2}{r}\, F_{0r}$ and for
that in (\ref{delta}) having a  delta function singularity. And
since $A_r=A'_r=0$, we conclude that all higher covariant
derivatives of the field strength coincide for both
configurations. This is precisely the condition ensuring that
there exist a gauge transformation connecting   $A_{\mu}$ and
$A'_{\mu}$ \cite{DD}-\cite{maj}. So that we conclude that
(\ref{dirit}) is gauge-equivalent to the original gauge field
configuration one gets in the $q \to \infty$ limit.

Then, we can write the resulting BPS equation for the $U_*(2)$
Yang-Mills-Higgs system and its monopole solution in the form
\be
\frac{1}{2} \varepsilon^{ijk}G_{jk} = D^i\Phi
\label{alfinbps}
\ee
\begin{eqnarray}
&&\vec \Phi = -\frac{1}{2r}\, \vec \Omega \; , \;\;\; \vec B_r   =
0\; , \;\; \vec B_\vartheta = -  \vec \Omega
\wedge \partial_\vartheta \vec \Omega \; , \nonumber \\
&& \vec B_\varphi = - \vec \Omega \wedge \partial_\varphi\vec
\Omega + \frac{1}{2} (1 + \cos \vartheta) \vec \Omega \cr
&&\Phi^4  = -\frac{1}{2r} \; , \;\; {B}_r^4  = 0  \; , \;\;
{B}_\vartheta^4 = 0  \; , \;\; {B}_\varphi^4 = -\frac{1}{2}(1 + \cos
\vartheta) \label{diritnueva}
\end{eqnarray}

With this time-independent configuration we can make the
correspondence (\ref{Higgs}) and obtain a BPS monopole. Note that
both the $SU(2)$ and $U(1)$ components of $B_\varphi$ have a
contribution $1/2(1 + \cos \vartheta)$ which coincide with the
Wu-Yang and Dirac singular monopole configuration. In order to
compute the corresponding magnetic charge, we define, as usual, an
``electromagnetic'' field strength ${{\cal G}_{ij}}$ by projecting
the $U_*(2)$ field strength along the $\Phi$ direction,

\be
{\cal G}_{ij} = {\rm tr}\left( \frac{\Phi}{|\Phi|}
G_{ij}\right)
\ee
which leads to a magnetic field of the form
\be
B^r = -\frac{1}{r^2}
\label{monop}
\ee
corresponding to a unit charge magnetic monopole
\be Q_m = \frac{1}{4\pi}\Phi_m = -1 \ee
with $\Phi_m$ the magnetic flux associated to (\ref{monop}). The
corresponding electric field,  consistently defined as
\be
{\cal G}_{i0} = {\rm tr}\left( \frac{\Phi}{|\Phi|}
G_{i0}\right)
\ee
of course vanishes. So, we have arrived to a magnetic
monopole-like solution of first order (BPS) equations
\be
D_i\Phi = \frac{1}{2} \varepsilon_{ijk}G_{jk}
\label{bps}
\ee
which are those giving the extrema for the energy of a gauge
field-Higgs system. Then, apart from the fact that there is a
Dirac-Wu-Yang singularity, the configuration solves the second
order Yang-Mills-Higgs equations of motion,
\begin{eqnarray}
D_i G^{ij} &=& [\Phi,D^j\Phi] \nonumber\\
D_iD^i \Phi &=& 0
\end{eqnarray}

Of course,  the energy associated to the solution
(\ref{diritnueva}), \be E =  {\rm Tr} \int d^3x \left( D_i\Phi
D_i\Phi + \frac{1}{2}F_{ij}
 F_{ij} \right) \label{sees} \ee
is strictly infinite (as it coincides with the selfenergy of a
Dirac monopole)
\be
E = \pi  \int dr \frac{1}{r^2} =  \int d^3 x B_{mon}^2
\ee
Now, if
we introduce a regulator $\epsilon$\footnote{Regulator $\epsilon$
 is dimensionless
since $r$ is a dimensionless variable.} to cut off the short-distances
divergence and
recover the dimensional scale $R$ ($\theta =\theta_0 R^2 = 2 R^2$) we
can write $E$ in the form
\be
E =  \frac{\pi}{g_{YM}^2 R\epsilon} =  \frac{\pi R }{g_{YM}^2 R^2 \epsilon} =
\frac{2\pi}{g_{YM}^2\theta} \frac{R}{\epsilon}
\label{E}
\ee
(We have reintroduced the gauge coupling constant $g_{YM}$ which was taken
 equal to 1 along the paper). Defining a length $L = R/\epsilon$
we see that $E$ can be identified with the mass of a string of
length $L$  whose tension is
\be
T = \frac{2\pi}{g_{YM}^2\theta}
\label{T} \ee
One can see (\ref{sees}) as emerging in the decoupling linearized
limit of a $D3$-brane in the Type IIB string theory with the Higgs
field describing its fluctuations in a transverse
direction\footnote{We thank the referee for clarifying to us the
correct brane interpretation of the solution.}. Since the
$B$-field leading to our noncommutative setting is transverse to
the $D3$-brane surface, one can make an analysis similar to that
presented by Callan-Maldacena in \cite{CM} with the scalar field
describing a perpendicular spike. In this last investigation,
where the electric case is discussed, the string interpretation
corresponds to an $F$-string attached to a $D3$-brane. Our
magnetic case can be related to this by  an $S$-duality
transformation changing the $F1$ into a $D1$ string. Comparing the
tension of such a $D1$-string with the one resulting from our
solution (eq.(\ref{T})),
\be
T_{D1} = \frac{1}{2\pi \alpha'g_s} = \frac{2\pi}{g_{YM}^2\theta}
\ee
and  using $2\pi g_s = g^2_{YM}$ we see that quantization of the
magnetic monopole charge leads to a quantized value  for $\theta$
in string length units equal to $1$ for our charge-1 monopole,
$\theta/2\pi \alpha' = 1$.

\section*{Discussion}

We shall  summarize here our results and discuss the properties of
the noncommutative monopole solution we have found as compared
with previous constructions.

Previous investigations on noncommutative monopoles
\cite{Gross1}-\cite{Gross3} were ba\-sed in Nahm's construction in
ordinary space \cite{nahm}. These works  start from the ADHM
version of the noncommutative multi-instanton and for the $U_*(1)$
gauge group, lead to a BPS solution which has zero magnetic
charge.

The alternative route we have taken, parallels in noncommutative
space, the observation of Refs.\ \cite{manton}, by taking the
infinite charge limit of an axially symmetric (in time) instanton.
The resulting  configuration  solves the BPS equations for a
Yang-Mills-Higgs system with the original $A_0$ gauge field
component playing the role of the scalar field.

In both approaches -that of ref.\cite{Gross1}-\cite{Gross3} and
ours- one needs to start from a  multi-instanton configuration in
noncommutative 4-dimensional space. If one follows the Nahm
approach, one needs a noncommutative version of the ADHM solution
and this was presented in \cite{NS}. The noncommutative solution
corresponds to a self-dual $\theta^{\mu\nu}$ which means that the
noncommutative relations are reduced to the nontrivial  pair
$[x^1,x^2] = [x^3,x^4] = i\theta$. In contrast, the axially
symmetric instanton solution   corresponds to a noncommutative
relation of the form  $[r,t] = i\theta(r,t)$ \cite{Correa2}
(Covariance arguments force the condition $\theta(r,t) = r^2
\theta_0$).

When the 4-dimensional original problem is reduced to three
dimensions, these different commutation relations lead, of course,
to different noncommutative spaces. In particular, one could think
that in our construction, for which noncommutativity necessarily
involves time, static configurations  could  just be considered as
ordinary commutative ones. However
%%%%%%%%% AGREGADO %%%%%%%%%%%%%%%%%%%%%%%%%%%%%%%%%%%%%%%%%%%%%%%%%%%%%%%%
this configuration has a genuine noncommutative origin as a
descendent of the noncommutative instanton
(\ref{instanI})-(\ref{instanIV}). Moreover
%%%%%%%%%%%%%%%%%%%%%%%%%%%%%%%%%%%%%%%%%%%%%%%%%%%%%%%%%%%%%%%%%%%%%%%%%%%
since solitons are intended to play a role through nonperturbative
effects where all space-time variables come into play, their
noncommutative character manifests, as it happens for example when
one computes tension (\ref{T})  from the string-monopole mass
formula.

%%%%%%%%% AGREGADO %%%%%%%%%%%%%%%%%%%%%%%%%%%%%%%%%%%%%%%%%%%%%%%%%%%%%%%%
It is worthwhile to emphasize how well Manton's method works for
the noncommutative instanton (\ref{instanI})-(\ref{instanIV})
leading, as in ordinary space, to a time-independent configuration
satisfying the BPS equations. And also how different are the final
products: a 't Hooft-Polyakov monopole in ordinary space and a
Wu-Yang monopole in the present case.
%%%%%%%%%%%%%%%%%%%%%%%%%%%%%%%%%%%%%%%%%%%%%%%%%%%%%%%%%%%%%%%%%%%%%%%%%%%

An application to brane dynamics of noncommutative monopoles was
given in \cite{Gross3} for the case a  static BPS $U_*(1)$
solution obtained from an ADHM instanton. Now, the soliton
obtained from the ADHM noncommutative instanton has zero  magnetic
charge, a result that can be understood in terms of a system of a
magnetic monopole attached to a flux tube of opposite charge,
transverse to the noncommutative plane. In contrast, we have shown
that the charge of the solution we obtained is effectively 1.
Studying the second order equations of motion associated to our
BPS solution, we have seen that our soliton  corresponds to  a
Wu-Yang singular configuration: although it verifies exactly the
BPS first order equations, delta-function sources are needed in
the second order Euler-Lagrange equations.

Let us finally point a direction along which it would be
worthwhile to pursue our investigation. As already mentioned, the
reduction from selfdual to BPS equations could be performed with
the isometry $k_\mu$ not necessarily in the Euclidean time
direction. In particular, a different choice for $k^\mu$ leads in
ordinary space to monopoles on $H^3$, hyperbolic 3-spaces, as
defined in \cite{atiyah}. Instead of the noncommutative axially
symmetric (with axis in time) instantons we started from, one
should consider   axially symmetric  invariant noncommutative
instantons but in this case with ``axis'' in $R^2 \sim S^1 \subset
R^4$. The properties of the resulting monopoles in the
corresponding noncommutative space will change drastically and can
exhibit interesting features. We hope to come back to this problem
in the future.

\vspace{1cm}

%%%%%%%%%%%%%%%%%%%%%%%%%%%%%%%%%%%%%%%%%%%%%%%%%%%%%%%%%%%%%%%%%%%%%%%%%%%%%%%%%%%%
%%%%%%%%%%%%%%%%%%%%%%%%%%%%%%%%%%%%%%%%%%%%%%%%%%%%%%%%%%%%%%%%%%%%%%%%%%%%%%%%%%%%

\noindent\underline{Acknowledgements}: We wish to thank Carlos
Nu\~nez for helpful comments. This work  was partially supported
by UNLP, CICBA, CONICET, ANPCYT (PICT grant 03-05179) Argentina
and ECOS-Sud Argentina-France collaboration (grant A01E02). D.H.C
was partially supported by Fundaci\'on Antorchas.

%%%%%%%%%%%%%%%%%%%%%%%%%%%%%%%%%%%%%%%%%%%%%%%%%%%%%%%%%%%%%%%%%%%%%%%%%%%%%%%%%%%%
%%%%%%%%%%%%%%%%%%%%%%%%%%%%%%%%%%%%%%%%%%%%%%%%%%%%%%%%%%%%%%%%%%%%%%%%%%%%%%%%%%%%
%%%%%%%%%%%%%%%%%%%%%%%%%%%%%%%%%%%%%%%%%%%%%%%%%%%%%%%%%%%%%%%%%%%%%%%%%%%%%%%%%%%%
%%%%%%%%%%%%%%%%%%%%%%%%%%%%%%%%%%%%%%%%%%%%%%%%%%%%%%%%%%%%%%%%%%%%%%%%%%%%%%%%%%%%
%%%%%%%%%%%%%%%%%%%%%%%%%%%%%%%%%%%%%%%%%%%%%%%%%%%%%%%%%%%%%%%%%%%%%%%%%%%%%%%%%%%%
%

%%%%%%%%%%%%%%%%%%%%%%%%%%%%%%%%%%%%%%%%%%%%%%%%%%%%%%%%%%%%%%%%%

\end{document}